# Artificial Intelligence based Smart Doctor using Decision Tree Algorithm

Rida Sara Khan, Asad Ali Zardar, Zeeshan Bhatti

*Abstract* — *Artificial Intelligence (AI) has already made a huge impact on our current technological trends. Through AI developments, machines are now given power and intelligence to behave and work like human mind. In this research project, we propose and implement an AI based health physician system that would be able to interact with the patient, do the diagnosis and suggest quick remedy or treatment of their problem. A decision tree algorithm is implemented in order to follow a top down searching approach to identify and diagnose the problem and suggest a possible solution. The system uses a questionnaire based approach to query the user (patient) about various Symptoms, based on which a decision is made and a medicine is recommended.*

## I. INTRODUCTION

The ability to learn is one of the most fundamental attributes of intelligent behavior [1]. Health is a key issue in current times and everyone now seems to be too busy to consider their day to day health issues. This is where an online health services come into play. The use of Artificial Intelligence (AI) to make systems behave and work more like humans is gaining popularity. The idea of using an AI system to diagnose and provide remedy for their daily minor health issues, could save tons of time and money of visiting and waiting for doctor at clinics. Sometime patients are unable to reach at doctors or hospital, moreover occasionally it happens that patient is alone and can't walk or is unable to see a doctor. Therefore, in various situations it has become a dire need to have a personal medical assistant by your side at all times, and what better way to have that then as an artificially intelligent smart doctor application on your cell phone[2].

In this project, the scope is to involve artificial intelligence to assist any person by diagnosing and providing the treatment. This will be an AI based smart doctor application that would behave as a doctor. The android operating system is being used commonly today [3].

The idea is to develop an android app for users providing a natural interaction with a softbot to diagnose the health problems. By using this user has access to the doctor completely so he/she can review his health. The smart doctor will be driven through an AI agent based algorithm that would take the users precept and make a decision based on the knowledge base and ma-chine learning techniques[4][5].

## II. SIMILAR WORK

There are numerous application that we found on app store that use a similar approach in assisting patients and diagnosing their health issue. An app called "ZOCDOC" is a medical care and scheduling application with search facility about medical practices and doctors [5]. It more of a personal health diary to keep your medical record, rather that diagnostics service. Figure .1 shows the basic user interface of the application [6].

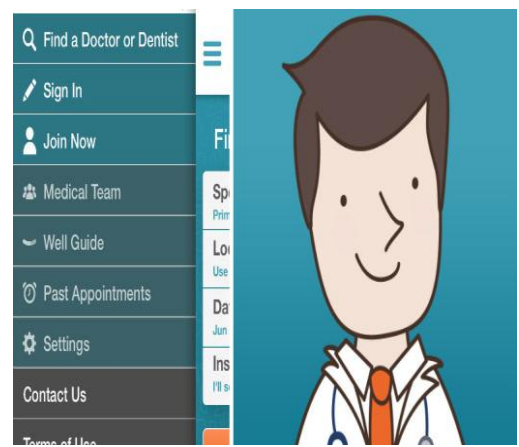

Figure 1: ZOCDOC app user interface

Similarly, another app called "Babylon's uses a AI technology to process various combination of symptoms at a very high accuracy and speed [8][7]. The system uses a chatbot to deal with "urgent but non-life-threatening conditions". Figure 2 shows the basic user interface of the application [9].

Manuscript Received: 21-6-2017; accepted: 2017; date of current version December- 2017
Rida Sara Khanis withInstitute of Information and Communication Technology, University of Sindh, Jamshoro, Pakistan (Email: rida_khan5@live.co.uk)
Asad Ali Zardar is with is withInstitute of Information and Communication Technology, University of Sindh, Jamshoro, Pakistan (email: zardariasadali@gmail.com)
Zeeshan Bhatti is withInstitute of Information and Communication Technology, University of Sindh, Jamshoro, Pakistan (Email: zeeshan.bhatti@usindh.edu.pk)

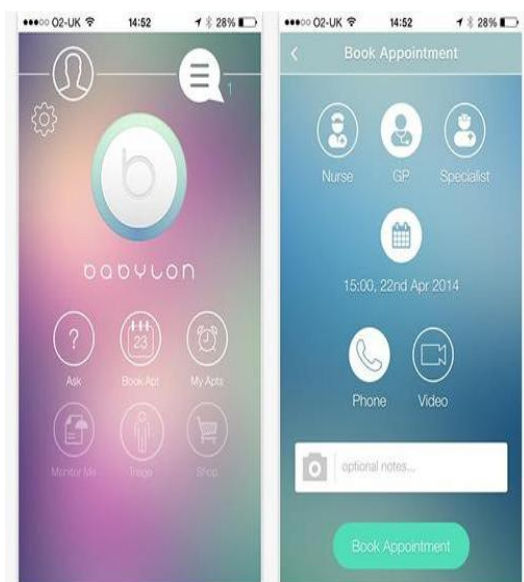

Figure 2: Babylon user interface

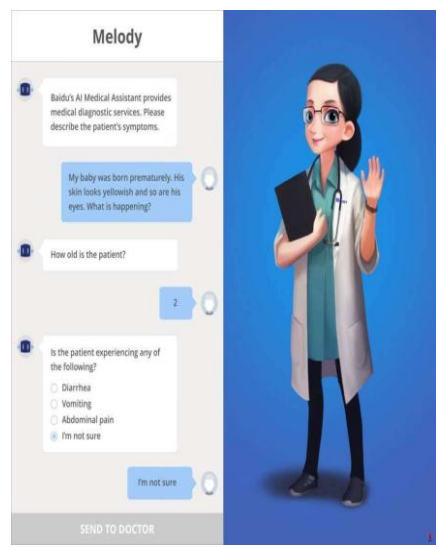

Figure 4: BAIDU user interface

Whereas, another app named "Pocket Doctor" is a learning and educational app that only provides information regarding the Health & Medicine, human anatomy, body max index calculations and many more similar features [10]. It also provided a health dictionary and based on which various symptoms can be analyzed [11][12][13].

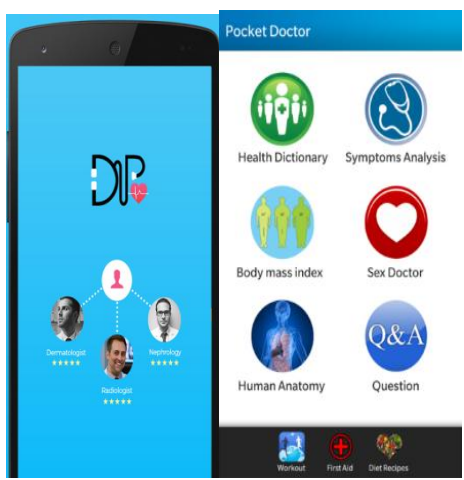

Figure 3: Pocket Doctor User interface

Similarly one of the most common medical app is "BAIDU" that uses an artificial intelligence with advance deep learning and natural language processing technologies based chatbot system called 'Melody', to facilitate patients and doctors. The figure 4 shows the BAIDU based chatbot called Melody [14][15].

### III. METHODOLOGY

The smart doctor project development involves using various key components of artificial intelligence such as natural language processing, smart agents, machine learning and knowledge base. A Rapid Application Development (RAD) software development model is used to design and develop the smart doctor app, with AI based chatbot system for communicating with the patient. The RAD model used in the system is illustrated in figure 5.

The system requirement phase involved identifying 3 key features needed to be incorporated into the application. First the system must understand what diseases patient have. Second the system must tell appropriate medicine or treatment to patient based on the precept of diagnostics from various questions through soft-bot. finally the system must set Alarm for patient to take medicine on time.

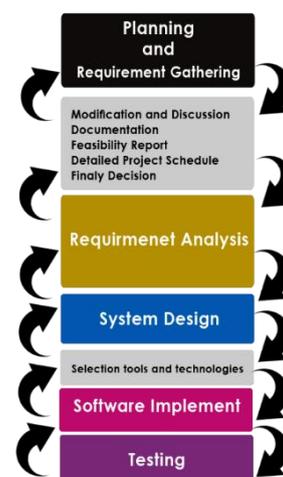

Figure 5: Rapid Application Model for Smart Doctor

Initially the system focuses on general health issues and their treatments, such as health problem involving general physician attaining to Ears, Nose Throat, Fever, Headaches, and Stomachache etc. The System design phase involves development knowledge base of various diseases and their symptoms, causes, treatments etc.

The AI based system architecture is shown in figure 6.

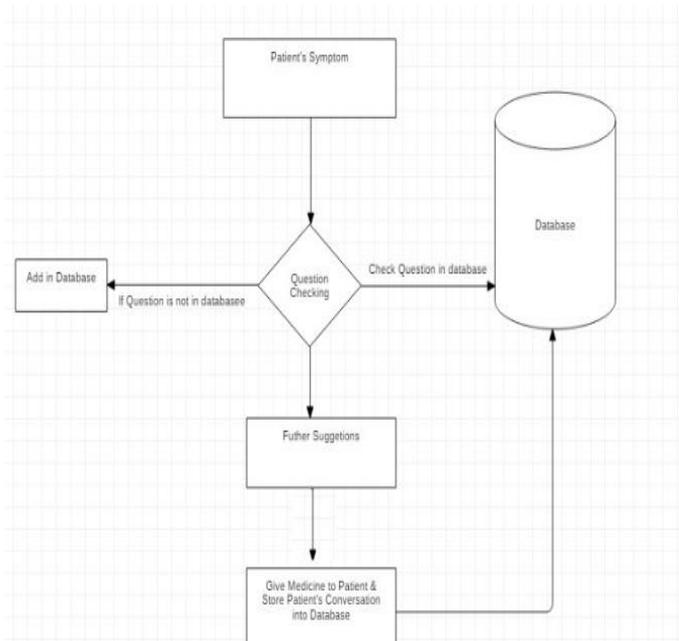

Figure 6: System Architecture

The AI system uses a rational agent driven through soft-bot with goal based agent to diagnose and suggest health treatment. The precept given from the user input would be used to provide suggestion that is then searched from a knowledge base using tree search algorithm.

The decision tree search algorithm used in our project drives the perception based AI softbot questionnaire, as illustrated in figure 7. We get new perception at every input from the user, which determines the next level of question, therefore the use of tree search algorithms facilitates the system through match strategies (input), if there is no match, and then the system would continue the loop until end of tree is reached.

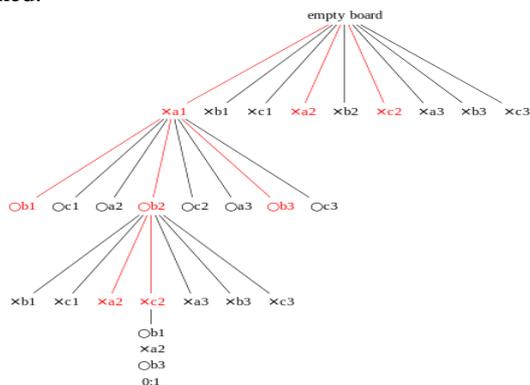

Figure 7: Tree search algorithm structure used in Smart doctor app

Based on tree search, the smart doctor uses a question and answer approach as shown in figure 8. In this system, each question is mapped with an id and then cross-referenced with symptoms provided by the user input and then again based on this a new question is formed and asked for the user.

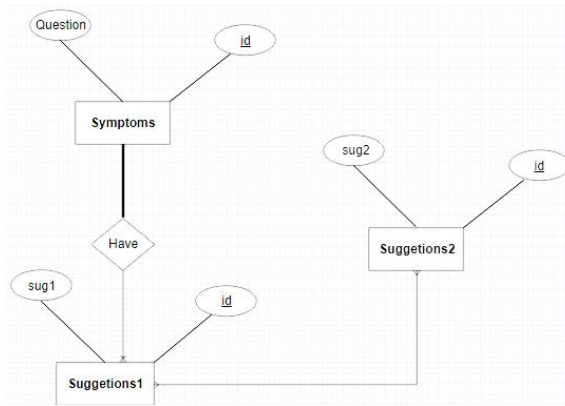

Figure 8: Tree search model of the system.

The decision algorithm used for simple disease uses a basic tree structure with minimum nodes as shown in figure 9.

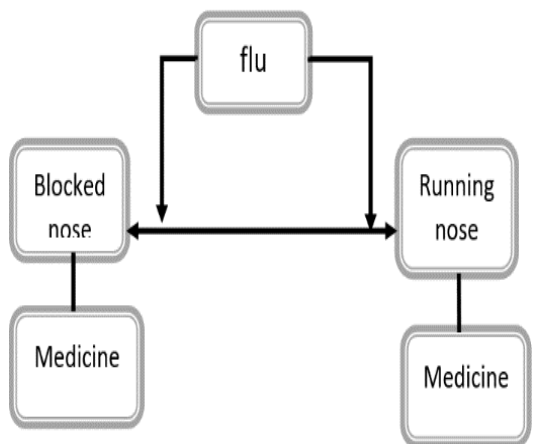

Figure 9: Decision tree for simple disease

The decision tree logic for a complex disease with more symptoms is created with more complex node structure based on multiple user quires and responses as illustrated in Figure 10.

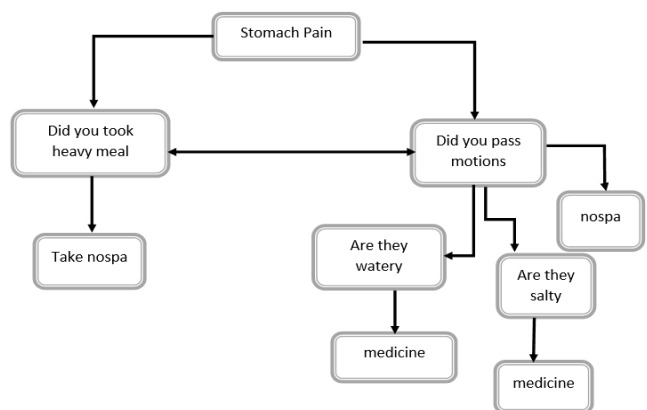

Figure 10: Decision tree for complex structure

This softbot based question and answer is better illustrated in the use case diagram, shown in figure 9.

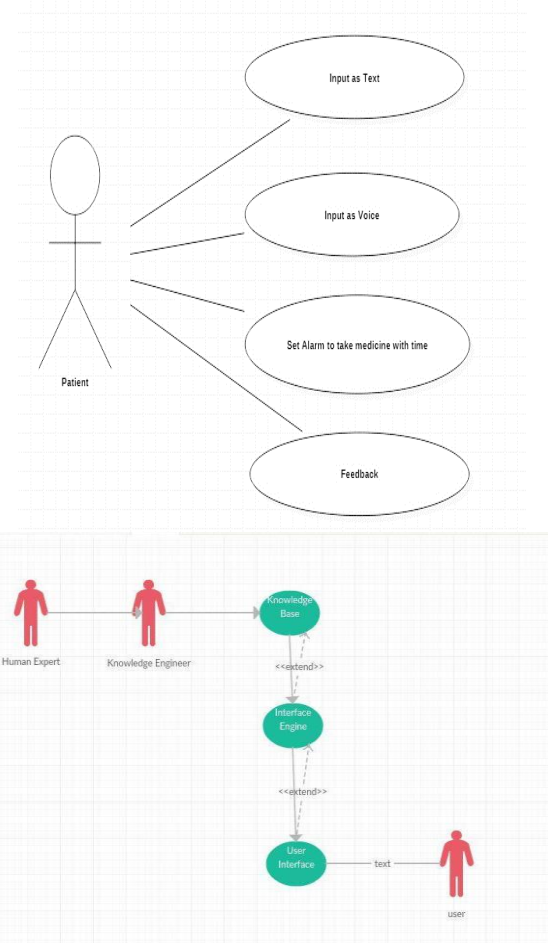

Figure 11: Use case diagram of the smart doctor system.

## IV. RESULT AND DISCUSSION

The final system of the smart doctor application is a working prototype using more than 100 diseases and their diagnostics at several layers of tree search. Table 1 shows few sample questions asked at initial level by the chatbot. Once the user replies to the replies to the first layer of questions, then the system again; based on user inputs, asks level two of questions as given in Table 2. The user again replies to the send level questions and the system then finally provides possible suggestions as described in Table 3.

Table 1: Symptoms asked by patient from Smart Doctor

| Serial No | Symptoms asked by Patient |
|---|---|
| 1 | I have pain in my neck |
| 2 | I have pain in my stomach |
| 3 | I have got sore throat |

Table 2: Suggestions given by System to the Patients

| Serial No | Suggestions given by System |
|---|---|
| 1 | Do you have vomiting too |
| 2 | Do you have diarchic |
| 3 | Do you ear-ache |

Table 3: Medicines given by system to patients

| Serial No | Medicine Given by System |
|---|---|
| 1 | You have migraine pain and I prescribe you to take Desprine and Bruefen and cream for massage. |
| 2 | I prescribe to take Flagel and avoid heavy junk food |
| 3 | I prescribe you to take Arimic , if taken then take Augmentin. |

The final app is working efficiently and as perceived. The app was test several times and has been checked by professional medical doctors who used it for verification. The Figure 10 shows the first startup screen of the app.

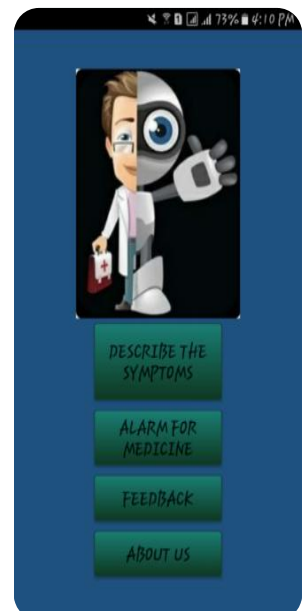

Figure 12: Startup screen of the smart doctor

The Figure.11 shows the interface of chatbot that is used for communication between the user and the AI agent. Through this user interface the AI agent asks quires and the user responds to those quires and finally a remedy is proposed by the system.

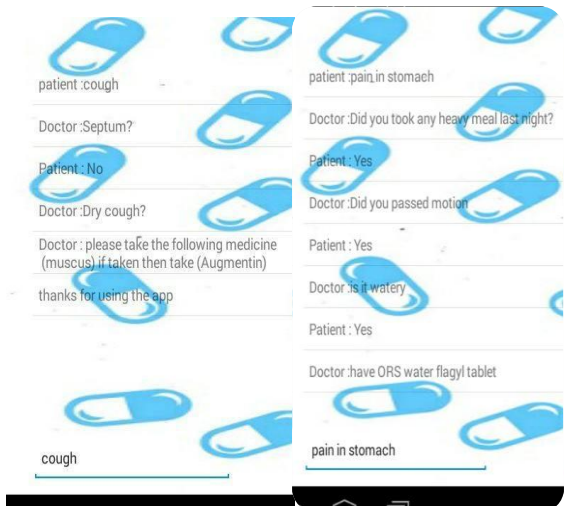

Figure 13: User interface of the Smart doctor softbot.

## V. CONCLUSION AND FUTURE WORK

Artificial Intelligence is becoming a norm of every technological development. Through this project the gap between Ai and health science has been narrowed down and a new possibilities have been introduced by which AI can be used to facilitate human development. The project uses a AI system to develop an efficient chatbot system based on rational agent with tree searching algorithm and knowledge base of medical data. The chatbot asks sever questions from the user or patient and then based on the replies suggests a possible remedy. The app was evaluat3ed and checked by several professional medical doctors and was considered accurate.

In future, the system need to increase its database and machine learning capabilities for much improved diagnostics. The use voice as input would also be an effective feature. The system needs to go to deeper level of tree from 3 to 7 and be more accurate.

ACKNOWLEDGEMENT

This final year project work was carried out in Multimedia Animation and Graphics (MAGic) Research Group under the supervision of Dr. Zeeshan Bhatti.

REFERENCES

[1] Michalski, R. S., Carbonell, J. G., & Mitchell, T. M. (Eds.)," *Machine learning: An artificial intelligence approach*" Springer Science & Business Media, pp.92, ,2013.
[2] Oliver Kharraz, Nick Ganju, Cyrus Massoumi , "Zocdoc: Find a Doctor – Doctor Reviews & Ratings"Date retrieved: 2017,pp.23-25, August 2017 . URL:https://www.zocdoc.com
[3] Babylon, "Babylon Health: Online Doctor Consultations & Advice" Retrived on 2017, pp. 34-38, 2017.URL:https://www.babylonhealth.com.
[4] Khoury Consulting, "Doctor Pocket" Retrieved on August-2017,vol.8(2):76-79, 2017, URL https://play.google.com/store/apps/details?id=com.docpoc.doctor&hl=en
[5] Baidu Doctor App, "Baidu's - Medical Robot", Retrieved on July 2017.
[6] S.M. Monzurur Rahman and Xinghuo Yu, "An unsupervised neural network approach to predictive data mining", Int. J. Data Mining, Modelling and Management, Vol. 3, No. 1,pp 18-41, 2011.
[7] Lev V. Utkin,"Regression analysis using the imprecise Bayesian normal mode", International Journal of Data Analysis Techniques and Strategies,Vol. 2, No.4 pp 356 – 372, 2010.
[8] Z.A. Al-Hemyari and I.H. Hussain, "Improved shrinkage testimators for the simple linear regression model", International Journal of Information and Decision Sciences, Vol. 4, No.1 pp 87 – 103, 2012.
[9] E. Heierman, III, and D. Cook, "Improving home automation by discovering regularly occurring device usage patterns", in Proc. 3rd IEEE Int. Conf. Data Mining, pp. 537–540, 2003.
[10] M. Ruotsalainen, T. Ala-Kleemola, and A. Visa, "Gais: A method for detecting interleaved sequential patterns from imperfect data,'' in Proc. IEEE Symp. Comput. Intell. Data Mining, pp. 530–534, 2007.
[11] Abel, David, James MacGlashan, and Michael L. Littman, "Reinforcement Learning As a Framework for Ethical Decision Making". Workshops at the Thirtieth AAAI Conference on Artificial Intelligence, 2016.
[12] Cserna, Bence, et al.,"Anytime versus Real-Time Heuristic Search for On-Line Planning". Ninth Annual Symposium on Combinatorial Search, 2016.
[13] Hamet, Pavel, and Johanne Tremblay, "Artificial Intelligence in Medicine". Metabolism, pp.10-15, 2017.
[14] Forestier, Germain, et al. ,"Automatic matching of surgeries to predict surgeons' next actions", Artificial Intelligence in Medicine,pp.12-15,2017.
[15] Jha, Saurabh, and Eric J. Topol, "Adapting to artificial intelligence: radiologists and pathologists as information specialists". JAMA 316.22, pp.2353-2354, 2